\pdfoutput=1
\documentclass[twoside,11pt]{article}
\PassOptionsToPackage{longnamesfirst,sort}{natbib}
\usepackage{obs_study_style}

\usepackage[T1]{fontenc}
\usepackage[utf8]{inputenc}
\usepackage{lmodern}
\usepackage{amsmath}
\usepackage{booktabs,tabularx,array}
\usepackage{listings}
\usepackage{microtype}
\usepackage{placeins}
\usepackage{xurl}
\usepackage[hidelinks]{hyperref}
\hypersetup{pdftitle={confoundvis: A Common Interface for Visualizing and Reporting Sensitivity Analyses for Unmeasured Confounding in R},pdfauthor={Subir Hait}}
\lstdefinestyle{Rout}{language={},basicstyle=\ttfamily\footnotesize,
  frame=leftline,framerule=0.6pt,framesep=6pt,xleftmargin=10pt,
  aboveskip=4pt,belowskip=9pt}

\newcommand{\fn}[1]{\texttt{#1}}
\let\oldus\_
\renewcommand{\_}{\oldus\allowbreak}
\newcommand{\cmark}{$\checkmark$}

\makeatletter
\def\ps@arxivplain{%
  \let\@mkboth\@gobbletwo
  \def\@oddhead{\hfill\small\thepage}%
  \def\@evenhead{\hfill\small\thepage}%
  \def\@oddfoot{}%
  \def\@evenfoot{}%
}
\makeatother
\firstpageno{1}

\title{confoundvis: A Common Interface for Visualizing and Reporting Sensitivity Analyses for Unmeasured Confounding in R}
\author{\name Subir Hait \email haitsubi@msu.edu \\
       \addr Measurement and Quantitative Methods\\
       College of Education, Michigan State University\\
       East Lansing, Michigan, USA}

\begin{document}
\raggedbottom
\shortcites{connors1996,frank2021,frank2025,werfel1998}
\maketitle

\begin{abstract}
Sensitivity analysis asks how strong unmeasured confounding would have to be before an observational conclusion changes. Several R packages already calculate established sensitivity quantities, but their outputs use different parameters, objects, and plotting conventions. In this article, we describe \texttt{confoundvis} 0.3.0, an R package designed as a common interface for these analyses. The package stores results from the impact threshold of a confounding variable, the robustness of inference to replacement, partial-$R^2$ omitted-variable bias analysis, E-values, and Rosenbaum bounds for matched pairs in one \texttt{confoundsens} object. Each result keeps its own parameter and its own published formula. The same functions then draw robustness curves, locate the points at which a conclusion changes, write plain-language reports, and collect several analyses in one table. When a framework describes a confounder by two associations, the package also draws the two-parameter surface on which the familiar single number lies and, where such benchmarks are defined, places measured covariates on the surface. We demonstrate the package with the right heart catheterization data and with a matched-pair study of welders. The native calculations agree with \texttt{sensemakr}, \texttt{konfound}, \texttt{DOS2}, \texttt{sensitivity2x2xk}, and \texttt{sensitivitymv} to numerical or grid precision, and imported \texttt{EValue} results reproduce the reported E-values. The purpose of \texttt{confoundvis} is not to replace the underlying methods or to place their sensitivity parameters on a common scale. Its purpose is to provide a consistent workflow for drawing, comparing, and reporting the results that those methods produce.
\end{abstract}


\section{Introduction}

Researchers who use observational data cannot be certain that every relevant confounder has been measured \citep{hernan2020,rosenbaum2002}. Even after observed covariates are controlled, an omitted factor may remain related to both the treatment and the outcome \citep{frank2000,rosenbaum1983}. Sensitivity analysis responds to this problem by asking how strong the omitted factor would have to be before the substantive or inferential conclusion changes. It does not identify the hidden variable, and it does not prove that hidden confounding is absent.

Several established methods translate this question into quantities that researchers can interpret. The impact threshold of a confounding variable (ITCV) of \citet{frank2000} uses the product of the omitted variable's correlations with the treatment and the outcome. The robustness of inference to replacement (RIR) of \citet{frank2013} asks how many cases would have to be replaced with cases in which the treatment has no effect. The partial-$R^2$ framework of \citet{cinelli2020} expresses omitted-variable bias through residual associations and provides robustness values. The E-value of \citet{vanderweele2017} expresses a related question on the risk-ratio scale. For matched designs, the sensitivity model of \citet{rosenbaum2002} bounds the odds of treatment within a matched pair by a factor $\Gamma$. These approaches address the same broad concern, but they do not use the same sensitivity parameter or the same decision rule.

Software for these methods is already available. The \texttt{sensemakr} package provides partial-$R^2$ calculations, formal observed-covariate bounds, and sensitivity plots \citep{sensemakrpkg,cinelli2024}. The \texttt{konfound} package provides the ITCV and the RIR \citep{xu2019,narvaiz2024}. The \texttt{EValue} package calculates E-values \citep{mathur2018}, and the \texttt{DOS2}, \texttt{sensitivity2x2xk}, and \texttt{sensitivitymv} packages provide Rosenbaum bounds \citep{dos2pkg,s2x2xkpkg,smvpkg,rosenbaum2015}. Therefore, the purpose of another software package should not be to reproduce these packages under a new name.

The practical problem addressed by \texttt{confoundvis} is narrower. Researchers may calculate a sensitivity analysis directly from a fitted model, receive a result from another package, or report several analyses in one study. They then need to see how the estimate changes, identify where a declared boundary is reached, compare a threshold with measured covariates when such a comparison is defined, and produce a clear report. These tasks require different object structures and plotting commands in each package. A common interface can reduce that friction while leaving the statistical meaning of each framework unchanged.

The above discussion implies that the main contribution of \texttt{confoundvis} \citep{hait2026} is its workflow. The package stores any one-dimensional sensitivity analysis as a path, that is, a grid of confounder strengths and the adjusted result at each strength, in a common S3 object called \texttt{confoundsens}. The same plotting, boundary-finding, reporting, and table functions can then be used whether the path was computed from a fitted model, imported from another package, or supplied by the user. Version 0.3.0 extends this workflow in three ways. First, the RIR and Rosenbaum bounds for matched pairs join the partial-$R^2$, ITCV, and E-value frameworks. Second, the package draws the two-parameter surface of each framework that defines one, so that the reader can see where the familiar single number lies among all confounders with the same effect. Third, one table collects several analyses, each in its own units. However, the package does not convert these quantities into one common score.

Given these considerations, the purpose of the current article is threefold. The first objective is to state the quantities that \texttt{confoundvis} computes or imports and the rules it uses to locate decision boundaries. The second objective is to explain the object structure and main functions of version 0.3.0. The third objective is to demonstrate the workflow with real observational data and to show that the native calculations agree with the established implementations. The article therefore focuses on how the software works rather than proposing a new sensitivity method.

The remainder of the article is organized as follows. Section~\ref{sec:quantities} reviews the quantities used by the package. Section~\ref{sec:design} describes the object and the functions. Section~\ref{sec:example} works through the right heart catheterization example, and Section~\ref{sec:pairs} works through a matched-pair example. Section~\ref{sec:validation} reports numerical validation, and Section~\ref{sec:discussion} discusses scope and limitations.

\section{Sensitivity quantities used by the package}
\label{sec:quantities}

The notation below is limited to what is needed to describe the software. The statistical derivations are given in the original methodological articles: \citet{frank2000} for the impact threshold of a confounding variable (ITCV); \citet{frank2013} and \citet{frank2021} for robustness of inference to replacement (RIR); \citet{cinelli2020} for partial-$R^2$ omitted-variable bias analysis; \citet{ding2016} and \citet{vanderweele2017} for the risk-ratio bias factor and E-value; and \citet{rosenbaum2002} and \citet{rosenbaum2009} for matched-pair sensitivity analysis and its amplification.

Let $D$ denote the treatment, $Y$ the outcome, $X$ the vector of measured covariates, and $U$ a hypothetical omitted confounder. For the regression-based analyses, consider the fitted ordinary least-squares model
\begin{equation}
Y_i=\gamma_0+\tau D_i+X_i^{\mathsf T}\beta+\varepsilon_i,\qquad i=1,\ldots,n.
\label{eq:ols}
\end{equation}
The fitted model contains $D$ and the measured covariates $X$, but not the hypothetical confounder $U$. Let $\widehat\tau$ denote the ordinary least-squares coefficient of $D$, $s=\operatorname{SE}(\widehat\tau)$ its conventional standard error, $t=\widehat\tau/s$ its $t$ statistic, and $d$ the residual degrees of freedom. These quantities provide the starting point for the regression-based sensitivity analyses described below.

\subsection{\texorpdfstring{Partial-$R^2$}{Partial-R2}}

Let $a=R^2_{D\sim U\mid X}$ and $b=R^2_{Y\sim U\mid D,X}$. Following \citet{cinelli2020}, the size of the omitted-variable bias is $B(a,b)=s\sqrt{d}\sqrt{ab/(1-a)}$, and the adjusted standard error is
\begin{equation}
\widetilde s(a,b)=s\sqrt{\frac{1-b}{1-a}}\sqrt{\frac{d}{d-1}}.
\label{eq:se}
\end{equation}
\fn{sens\_path\_lm()} evaluates the adverse path $\widetilde\tau=\operatorname{sign}(\widehat\tau)(|\widehat\tau|-B)$ on a grid of $\lambda=a$ values with $b=\lambda$. Its point-null crossing is the robustness value
\begin{equation}
RV=\tfrac12\left(\sqrt{f^4+4f^2}-f^2\right),\qquad f=|t|/\sqrt d .
\label{eq:rv}
\end{equation}
The significance crossing uses the adjusted standard error in Equation~\ref{eq:se} and the $t$ distribution with $d-1$ degrees of freedom.

\subsection{Impact threshold}

The observed partial correlation corresponding to the treatment coefficient can be written as $r=t/\sqrt{t^2+d}$, using the one-to-one relationship between a regression $t$ statistic and its corresponding partial correlation \citep{frank2000,frank2025}. At two-sided level $\alpha$, let $t_c$ denote the critical $t$ value with $d$ residual degrees of freedom and define $r_c=t_c/\sqrt{t_c^2+d}$. For a statistically significant result, \fn{itcv\_lm()} returns the conditional impact threshold of \citet{frank2000},
\begin{equation}
ITCV=\frac{|r|-r_c}{1-r_c},
\label{eq:itcv}
\end{equation}
together with the corresponding unconditional threshold
\begin{equation}
ITCV_{\mathrm{uncond}}=ITCV\sqrt{(1-R^2_{D\mid X})(1-R^2_{Y\mid X})},
\label{eq:itcv_uncond}
\end{equation}
which expresses the threshold on the zero-order correlation scale \citep{frank2000,frank2025}.

More generally, if a hypothetical confounder $U$ has partial correlations $r_{DU}$ with the treatment and $r_{YU}$ with the outcome, then the treatment-outcome partial correlation after adjusting for $U$ is
\begin{equation}
r_{\mathrm{adj}}=\frac{r-r_{DU}\,r_{YU}}{\sqrt{(1-r_{DU}^2)(1-r_{YU}^2)}},
\label{eq:radj}
\end{equation}
the recursive partial-correlation identity used in the ITCV development \citep{xu2019,frank2025}. For a fixed product of the two confounder correlations, the smallest product capable of crossing a specified inferential threshold occurs under the equal-strength configuration in the adverse direction \citep{xu2019,frank2025}. Therefore, the one-dimensional ITCV path uses $\lambda=k$, with $|r_{DU}|=|r_{YU}|=\sqrt{k}$, so that $r_{\mathrm{adj}}=\operatorname{sign}(r)(|r|-k)/(1-k)$. Along this path the adjusted correlation reaches $r_c$ at $k=ITCV$ and reaches zero at $k=|r|$.

\subsection{Robustness of inference to replacement}

Let $\beta^{\#}=\operatorname{sign}(\widehat\tau)\,t_c\,s$, with $t_c$ the critical value on $d$ degrees of freedom, be the smallest estimate in absolute value that would still be significant, holding the standard error fixed as in \texttt{konfound}. \citet{frank2013} define the share of the estimate that would have to be due to bias as $\pi=1-\beta^{\#}/\widehat\tau$, and the RIR as $\pi$ times the sample size, rounded to the nearest case; the case-replacement interpretation is further developed by \citet{frank2021}. \fn{rir\_lm()} computes these quantities and stores the path $\theta(\lambda)=(1-\lambda)\widehat\tau$, the estimate after a proportion $\lambda$ of the cases is replaced with cases in which the effect is zero. The path reaches $\beta^{\#}$ at $\lambda=\pi$. A substantive threshold can be supplied in place of $\beta^{\#}$, as recommended by \citet{frank2025}.

\subsection{E-value}

For a risk ratio $RR\ge1$, the E-value is $RR+\sqrt{RR(RR-1)}$ \citep{vanderweele2017}. When the confounder's risk ratios with the treatment and the outcome are $a$ and $b$ (in this subsection), the bias factor is
\begin{equation}
B_{\mathrm{RR}}(a,b)=\frac{ab}{a+b-1}
\label{eq:bf}
\end{equation}
\citep{ding2016}. \fn{from\_evalue()} stores $\theta(g)=\log RR-\log B_{\mathrm{RR}}(g,g)$, which crosses zero at the E-value. When the source result includes a confidence interval, the adapter reconstructs a log-scale standard error from the interval and holds it constant along the path. The path is a bound, not an identified adjusted risk ratio.

\subsection{Rosenbaum bounds for matched pairs}

In a matched-pair study, Rosenbaum's model allows the two units in a pair to differ in their odds of treatment by at most a factor $\Gamma\geq1$ \citep{rosenbaum2002}. For each $\Gamma$, the one-sided P-value of a randomization test has a sharp upper bound. \fn{sens\_path\_pairs()} computes this bound for Wilcoxon's signed rank statistic with the large-sample approximation used by \texttt{DOS2::senWilcox()}, or, for a binary outcome, the exact binomial bound for McNemar's statistic. It stores the bound as an equivalent standard normal deviate, so that the generic functions apply unchanged. The sensitivity value is the $\Gamma$ at which the bound reaches $\alpha$; the package finds it by root finding as well as on the grid.

\subsection{Two-parameter surfaces}
\label{sec:surfaces}

Most of the frameworks above describe a confounder by two associations, and a path is one line through the plane of those two associations. \fn{sens\_surface()} evaluates the framework's own formula over the whole plane and traces the boundaries at which the conclusion changes. For the partial-$R^2$ framework, the surface is the adjusted estimate over $(a,b)$, the plane drawn by \texttt{sensemakr}'s contour plot. The null boundary is $b=f^2(1-a)/a$, and the significance boundary is found numerically from Equation~\ref{eq:se}. The null boundary crosses the diagonal $a=b$ at the robustness value, and the significance boundary crosses it at the robustness value for significance. For the ITCV, the surface is Equation~\ref{eq:radj} over $(r_{DU},r_{YU})$, and the boundaries $|r_{\mathrm{adj}}|=r_c$ are solved in closed form. Pairs with $1-r^2-r_{DU}^2-r_{YU}^2+2r\,r_{DU}r_{YU}\le0$ cannot occur in any correlation matrix and are marked as infeasible. For E-values, the null boundary is the curve $B_{\mathrm{RR}}(a,b)=RR$, which passes through the point (E-value, E-value), as in the bias plot of \texttt{EValue}. For Rosenbaum's model, \citet{rosenbaum2009} show that $\Gamma=(\Lambda\Delta+1)/(\Lambda+\Delta)$, where $\Lambda$ bounds the association of $U$ with treatment and $\Delta$ bounds its association with the outcome. The significance boundary is the set of $(\Lambda,\Delta)$ pairs that give the sensitivity value, the pairs returned by \texttt{sensitivitymv::amplify()}. RIR has one parameter and therefore no surface.

The surface reports only these boundaries. The size of a shaded zone depends on the axis ranges chosen and carries no inferential meaning, so the package reports no area or share of the plane.

\subsection{Measured-covariate benchmarks}

For each measured term $Z$ in a fitted linear model, with $W$ denoting the other covariates, \fn{covariate\_impacts()} computes two reference quantities. The ITCV metric is the signed product $\operatorname{sign}(\widehat\tau)\,r_{DZ\mid W}\,r_{YZ\mid W}$, so that a positive value acts to reduce the estimate. The partial-$R^2$ metric is the bias index
\begin{equation}
I_Z=\sqrt{\frac{R^2_{Y\sim Z\mid D,W}\,R^2_{D\sim Z\mid W}}{1-R^2_{D\sim Z\mid W}}},
\label{eq:bias}
\end{equation}
which is compared with $f=|t|/\sqrt d$. The output also includes each term's zero-order correlations, their product (the zero-order impact), and, for terms whose zero-order impact acts against the estimate, the ratio of the unconditional ITCV to that impact, as in the practical guide of \citet{frank2025}. These quantities condition on $W$ rather than on all of $X$. A benchmark is a reference for discussion, not an upper bound on omitted confounding \citep{hosman2010,cinelli2020,lonati2026}.

\section{Package design}
\label{sec:design}

\subsection{The \texttt{confoundsens} object}

Every path is stored as an S3 object of class \texttt{confoundsens}. The constructor
\par\noindent\begin{minipage}{\linewidth}
\begin{lstlisting}
new_confoundsens(lambda, theta, level = NULL, se = NULL, t = NULL, meta = list())
\end{lstlisting}
\end{minipage}\par
checks the fields in Table~\ref{tab:object}. The numerical fields hold the path, and the \fn{meta} list records what the numbers mean. This structure does not make the underlying methods equivalent. Its role is computational. The other functions read \fn{meta} to choose axis labels, the null value, and the decision rule. Therefore, one function can draw a regression coefficient path, a correlation path, a log risk-ratio path, and a path of normal deviates without treating them as one scale.

\begin{table}[htbp]
\caption{Fields of a \texttt{confoundsens} object}
\label{tab:object}
\centering\small
\setlength{\tabcolsep}{4pt}
\renewcommand{\arraystretch}{1.15}
\begin{tabularx}{\textwidth}{@{}p{.15\linewidth}X@{}}
\toprule
\textbf{Field} & \textbf{Content} \\
\midrule
\fn{lambda} & Grid of sensitivity parameter values (partial $R^2$, correlation product, proportion of cases replaced, confounder risk ratio, or $\Gamma$). \\
\fn{theta} & Path values at each grid point (adjusted coefficient, adjusted partial correlation, estimate after replacement, log bias-adjusted risk ratio, or deviate of the upper-bound P-value). \\
\fn{se}, \fn{t} & Optional adjusted standard errors and test statistics along the path. \\
\fn{level} & Optional labels for several paths in one object, for example subgroups; plots facet by level, and crossings are found within each level. \\
\fn{meta} & List recording the framework, source, treatment, starting estimate, null value, decision threshold, axis labels, and framework summaries such as \fn{rv}, \fn{itcv}, \fn{rir}, or \fn{sens\_value}. \\
\bottomrule
\end{tabularx}
\end{table}

\subsection{Entry routes}

Table~\ref{tab:routes} lists the ways to create a path. The native functions start from a single-outcome model fitted by \fn{stats::lm()} or from a vector of matched-pair differences. The adapters read documented fields of the source objects and do not call the source packages. These packages are therefore suggested, not required.

\begin{table}[htbp]
\caption{Ways to create a \texttt{confoundsens} object}
\label{tab:routes}
\centering\small
\setlength{\tabcolsep}{4pt}
\renewcommand{\arraystretch}{1.15}
\begin{tabularx}{\textwidth}{@{}>{\raggedright\arraybackslash}p{.36\linewidth}>{\raggedright\arraybackslash}p{.15\linewidth}X@{}}
\toprule
\textbf{Function} & \textbf{Input} & \textbf{Path and stored decision information} \\
\midrule
\fn{sens\_path\_lm(model, treatment, lambda)} & \texttt{lm} fit & Partial-$R^2$ path with \fn{se}, \fn{t}, \fn{dof}, and \fn{rv}. \\
\fn{itcv\_lm(model, treatment, alpha, lambda)} & \texttt{lm} fit & List with \fn{r}, \fn{r\_crit}, \fn{itcv}, \fn{itcv\_unconditional}, benchmark ratios, and the path in \fn{\$path}. \\
\fn{rir\_lm(model, treatment, alpha, threshold)} & \texttt{lm} fit & List with $\beta^{\#}$, percent bias, RIR, and the path in \fn{\$path}. \\
\fn{sens\_path\_pairs(d, gamma, method)} & Pair differences & Rosenbaum path (Wilcoxon or binary) with \fn{sens\_value} and upper-bound P-values. \\
\fn{from\_sensemakr(x, lambda)} & \texttt{sensemakr} object & Partial-$R^2$ path; formal bounds attached as \fn{meta\$bounds}. \\
\fn{from\_konfound(x, dof, lambda, index)} & \texttt{konfound} raw output & ITCV path (\fn{index = "IT"}) or RIR path (\fn{index = "RIR"}). \\
\fn{from\_evalue(x, lambda)} & \texttt{evalue} object & Log risk-ratio path with a constant log-scale standard error when the interval is available. \\
\fn{new\_confoundsens()}, \fn{as\_confoundsens()} & Vectors or data frame & User-supplied path. \\
\bottomrule
\end{tabularx}
\end{table}

\subsection{Functions that operate on paths}

\fn{robustness\_points(x, alpha = 0.05)} finds, within each level, the smallest $\lambda$ at which the point estimate reaches the null and the smallest $\lambda$ at which significance is lost, by linear interpolation between grid points. The significance rule depends on what the object carries: a stored threshold (\fn{theta threshold}), adjusted $t$ values with stored degrees of freedom (\fn{t, df = d - 1}), or, failing these, a normal interval or normal $t$ rule. The rule used is returned in the \fn{sig\_rule} column of \fn{robustness\_points()} and \fn{sens\_table()}. The direction of the original conclusion is taken from the stored threshold, so negative estimates are handled correctly. \fn{sens\_report()} turns these results into a short text that ends with a statement of what the analysis cannot establish. \fn{sens\_table()} collects several paths in one data frame, one row per path, with the framework summary and the crossings in each path's own units.

The plotting functions return \texttt{ggplot2} objects \citep{wickham2016}. \fn{plot\_robustness\_curve()} draws a path with its null line, stored threshold, and pointwise bands when standard errors are available. \fn{plot\_sensitivity\_love()} draws the output of \fn{covariate\_impacts()} in the layout of a covariate-balance Love plot \citep{austin2009}. \fn{plot\_sensitivity\_contour()} draws the exact ITCV boundaries of Equation~\ref{eq:radj}, the product curve $|r_{DU}r_{YU}|=ITCV$ for reference, and the infeasible region. \fn{plot\_sensitivity\_surface()} draws the output of \fn{sens\_surface()} with benchmarks. Labels are separated with \texttt{ggrepel} when it is installed \citep{ggrepelpkg}.

Table~\ref{tab:capability} shows which functions apply to which framework. The curve, report, and table functions apply to every path. The surface, contour, and benchmark functions apply only where the framework defines them.

\begin{table}[htbp]
\caption{Common and framework-specific functions}
\label{tab:capability}
\centering\small
\setlength{\tabcolsep}{5pt}
\renewcommand{\arraystretch}{1.15}
\begin{tabularx}{\textwidth}{@{}>{\raggedright\arraybackslash}Xccccc@{}}
\toprule
\textbf{Function} & \textbf{Partial $R^2$} & \textbf{ITCV} & \textbf{RIR} & \textbf{E-value} & \textbf{Rosenbaum} \\
\midrule
\fn{robustness\_points()}, \fn{plot\_robustness\_curve()} & \cmark & \cmark & \cmark & \cmark & \cmark \\
\fn{sens\_report()}, \fn{sens\_table()} & \cmark & \cmark & \cmark & \cmark & \cmark \\
\fn{sens\_surface()}, \fn{plot\_sensitivity\_surface()} & \cmark & \cmark & -- & \cmark & \cmark \\
\fn{covariate\_impacts()}, \fn{plot\_sensitivity\_love()} & \cmark & \cmark & -- & -- & -- \\
\fn{plot\_sensitivity\_contour()} & -- & \cmark & -- & -- & -- \\
\bottomrule
\end{tabularx}
\end{table}

\subsection{Relationship to existing software}

Table~\ref{tab:software} summarizes the division of labor. The upstream packages remain the reference implementations of their methods. \texttt{confoundvis} reproduces their calculations, imports their results, and adds the common object and the functions that work across frameworks.

\begin{table}[htbp]
\caption{Division of labor between \texttt{confoundvis} and the packages it works with}
\label{tab:software}
\centering\small
\setlength{\tabcolsep}{4pt}
\renewcommand{\arraystretch}{1.15}
\begin{tabularx}{\textwidth}{@{}>{\raggedright\arraybackslash}p{.21\linewidth}p{.33\linewidth}X@{}}
\toprule
\textbf{Package} & \textbf{Native statistics and graphics} & \textbf{Role of \texttt{confoundvis}} \\
\midrule
\texttt{sensemakr} & Partial-$R^2$ statistics, robustness values, formal bounds; contour and extreme-scenario plots & \fn{sens\_path\_lm()} reproduces the path; \fn{from\_sensemakr()} imports the result and its bounds \\
\texttt{konfound} & ITCV and RIR; ITCV and RIR figures & \fn{itcv\_lm()} and \fn{rir\_lm()} reproduce the results; \fn{from\_konfound()} imports them \\
\texttt{EValue} & E-values for several effect measures; bias plots & \fn{from\_evalue()} converts the risk-ratio result \\
\texttt{DOS2}, \texttt{sensitivity2x2xk}, \texttt{sensitivitymv} & Rosenbaum bounds for many designs and statistics; amplification & \fn{sens\_path\_pairs()} reproduces the matched-pair Wilcoxon and McNemar bounds \\
\texttt{confoundvis} & Common object; curves, surfaces, contours, Love plots, boundaries, reports, tables & Integration and reporting layer; no new sensitivity measure \\
\bottomrule
\end{tabularx}
\end{table}

\FloatBarrier

\section{Worked example: right heart catheterization}
\label{sec:example}

\subsection{Data and starting model}

We use the right heart catheterization (RHC) data distributed with the \texttt{ATbounds} package \citep{lee2026}. The data come from the observational study of \citet{connors1996}, later analyzed by \citet{hirano2001}. The extract contains 5,735 critically ill patients, 2,184 of whom received catheterization, and 72 covariate columns. We define death within 30 days as one minus the recorded survival indicator. The example demonstrates how the package operates. It does not reproduce the original analysis or offer a new causal estimate.

The native functions start from an \texttt{lm} fit, so we fit a linear probability model with all covariates. The code below is the start of the replication script; every later chunk continues the same session.

\par\noindent\begin{minipage}{\linewidth}
\begin{lstlisting}
library(confoundvis)
options(width = 80, digits = 7)
data("RHC", package = "ATbounds")
rhc <- as.data.frame(RHC)
rhc$death <- 1 - rhc$survival
xvars <- setdiff(names(rhc), c("survival", "death"))
fit <- lm(reformulate(xvars, response = "death"), data = rhc)
summary(fit)$coefficients["RHC", ]
\end{lstlisting}
\end{minipage}\par
\par\noindent\begin{minipage}{\linewidth}
\begin{lstlisting}[style=Rout]
    Estimate   Std. Error      t value     Pr(>|t|)
6.336115e-02 1.341526e-02 4.723064e+00 2.379256e-06
\end{lstlisting}
\end{minipage}\par

All 5,735 patients enter the model, the design matrix has rank 74, and $d=5{,}661$. The adjusted RHC coefficient is 0.0634 with standard error 0.0134. Because the outcome is binary, these conventional standard errors are not the preferred inference for mortality, and a causal reading would need identification assumptions that this article does not examine. The package inherits the assumptions of the model the user supplies.

\subsection{\texorpdfstring{Partial-$R^2$ path}{Partial-R2 path}}

\fn{sens\_path\_lm()} builds the equal-strength path. A fine grid keeps the interpolated crossings close to their closed-form values.

\par\noindent\begin{minipage}{\linewidth}
\begin{lstlisting}
grid <- seq(0, 0.12, by = 0.0001)
p_r2 <- sens_path_lm(fit, treatment = "RHC", lambda = grid)
robustness_points(p_r2)
sens_report(p_r2)
\end{lstlisting}
\end{minipage}\par
\par\noindent\begin{minipage}{\linewidth}
\begin{lstlisting}[style=Rout]
   level theta_start lambda_null lambda_sig     sig_rule
1 pooled  0.06336115  0.06083428 0.03604828 t, df = 5660
Sensitivity report (framework: partial_r2; source: lm)
Treatment: RHC
Estimate 0.0634 (SE 0.0134, df 5661); robustness value RV = 0.0608.
Confounder strength (lambda): Confounder partial R2 with treatment and
  outcome.
Starting at 0.0634, the point estimate reaches the null at lambda =
  0.0608; significance (alpha = 0.05) is lost at lambda = 0.036.
These values describe how strong an omitted confounder would need to
  be. They do not show whether such a confounder exists, and they rest
  on the framework's own modelling assumptions. Judge plausibility
  against substantive knowledge and observed benchmarks (see
  covariate_impacts()).
\end{lstlisting}
\end{minipage}\par

An omitted variable that explained 6.08\% of the residual variance of both treatment and outcome would remove the point estimate. One that explained 3.60\% would remove significance at $\alpha=0.05$. The \fn{sig\_rule} entry records that the second value uses $d-1=5{,}660$ degrees of freedom.

\subsection{Impact threshold path}

The same model can be passed to \fn{itcv\_lm()}.

\par\noindent\begin{minipage}{\linewidth}
\begin{lstlisting}
it <- itcv_lm(fit, treatment = "RHC", alpha = 0.05, lambda = grid)
unlist(it[c("r", "r_crit", "itcv", "itcv_unconditional")])
robustness_points(it$path)
\end{lstlisting}
\end{minipage}\par
\par\noindent\begin{minipage}{\linewidth}
\begin{lstlisting}[style=Rout]
                 r             r_crit               itcv itcv_unconditional
        0.06265032         0.02604636         0.03758285         0.02877045
   level theta_start lambda_null lambda_sig        sig_rule
1 pooled  0.06265032  0.06265031 0.03758285 theta threshold
\end{lstlisting}
\end{minipage}\par

The significance crossing equals the conditional ITCV, and the null crossing equals $|r|$. An omitted variable whose partial correlations with RHC and with death each had magnitude $\sqrt{0.0376}=0.194$ in the adverse direction would make the association nonsignificant. This path uses $d$ degrees of freedom, following \citet{frank2000} and \texttt{konfound}, whereas the partial-$R^2$ path uses $d-1$. Each path keeps the convention of its own framework. For the partial-$R^2$ path, \fn{sig\_rule} reports the degrees of freedom; for the ITCV path, they are built into the stored critical correlation.

\subsection{Robustness of inference to replacement}

\fn{rir\_lm()} answers the RIR question for the same model.

\par\noindent\begin{minipage}{\linewidth}
\begin{lstlisting}
rr <- rir_lm(fit, treatment = "RHC")
unlist(rr[c("estimate", "beta_threshold", "perc_bias", "rir", "n_obs")])
sens_report(rr$path)
\end{lstlisting}
\end{minipage}\par
\par\noindent\begin{minipage}{\linewidth}
\begin{lstlisting}[style=Rout]
      estimate beta_threshold      perc_bias            rir          n_obs
  6.336115e-02   2.629906e-02   5.849340e+01   3.355000e+03   5.735000e+03
Sensitivity report (framework: rir; source: lm)
Treatment: RHC
Estimate 0.0634; threshold 0.0263. To nullify the inference, 58.49% of
  the estimate would have to be due to bias: replacing 3355 of 5735
  cases with cases for which the effect is zero (RIR = 3355).
Sensitivity parameter (lambda): Proportion of cases replaced with
  null-effect cases.
Starting at 0.0634, the estimate falls to the threshold when a
  proportion 0.585 of the cases is replaced.
RIR describes how many cases would have to be replaced with cases in
  which the treatment has no effect. It does not show that such cases
  exist; judge plausibility against knowledge of the sample and design.
\end{lstlisting}
\end{minipage}\par

To change the inference, 58.5\% of the estimate would have to be due to bias. Equivalently, 3,355 of the 5,735 patients would have to be replaced with patients for whom catheterization has no effect.

\subsection{Importing sensemakr and konfound results}
\label{sec:import}

A researcher who has already run \texttt{sensemakr} or \texttt{konfound} can convert the result instead of computing it again. Below, both packages are run on the same model, and the converted paths are compared with the native ones.

\par\noindent\begin{minipage}{\linewidth}
\begin{lstlisting}
s <- sensemakr::sensemakr(fit, treatment = "RHC",
                          benchmark_covariates = "surv2md1", kd = 1:3)
p_sm <- from_sensemakr(s, lambda = grid)
k_it <- konfound::konfound(fit, RHC, index = "IT", to_return = "raw_output")
p_kf <- from_konfound(k_it, dof = df.residual(fit), lambda = grid,
                      treatment = "RHC")
k_rir <- konfound::konfound(fit, RHC, index = "RIR", to_return = "raw_output")
p_kr <- from_konfound(k_rir, index = "RIR", treatment = "RHC")

c(sensemakr_theta = max(abs(p_sm$theta - p_r2$theta)),
  konfound_itcv   = max(abs(p_kf$theta - it$path$theta)),
  konfound_rir    = max(abs(p_kr$theta - rr$path$theta)),
  rir_cases       = k_rir$RIR_primary - rr$rir)
\end{lstlisting}
\end{minipage}\par
\par\noindent\begin{minipage}{\linewidth}
\begin{lstlisting}[style=Rout]
sensemakr_theta   konfound_itcv    konfound_rir       rir_cases
              0               0               0               0
\end{lstlisting}
\end{minipage}\par

The imported and native paths are identical at every grid point, and the two RIRs agree. The \texttt{sensemakr} object also carries formal bounds for a confounder one to three times as strong as \fn{surv2md1}, the two-month survival estimate. \fn{from\_sensemakr()} attaches these bounds to the path without changing them, and they appear later on the surface in Figure~\ref{fig:surface}.

\subsection{Importing an E-value result}

\texttt{confoundvis} does not fit risk-ratio models. For the risk-ratio analysis, we fit a modified Poisson model with the same covariates and HC0 standard errors \citep{zou2004,zeileis2004}, compute E-values with \texttt{EValue}, and convert the result.

\par\noindent\begin{minipage}{\linewidth}
\begin{lstlisting}
fit_rr <- glm(formula(fit), data = rhc, family = poisson(link = "log"))
V <- sandwich::vcovHC(fit_rr, type = "HC0")
beta <- unname(coef(fit_rr)["RHC"])
se_rr <- sqrt(V["RHC", "RHC"])
ci_rr <- exp(beta + c(-1, 1) * qnorm(0.975) * se_rr)
e <- EValue::evalues.RR(exp(beta), lo = ci_rr[1], hi = ci_rr[2])
p_ev <- from_evalue(e, lambda = seq(1, 2, by = 0.001))
e
robustness_points(p_ev)
\end{lstlisting}
\end{minipage}\par
\par\noindent\begin{minipage}{\linewidth}
\begin{lstlisting}[style=Rout]
            point    lower    upper
RR       1.102754 1.059514 1.147758
E-values 1.439372 1.310622       NA
   level theta_start lambda_null lambda_sig  sig_rule
1 pooled  0.09781046    1.439372   1.310622 normal CI
\end{lstlisting}
\end{minipage}\par

The path starts at $\log(1.1028)=0.0978$ because it is stored on the log scale. Its null crossing reproduces the point E-value (1.439). Because the Wald interval is symmetric on the log scale, its \fn{normal CI} crossing reproduces the confidence-limit E-value (1.311). For an interval that is not symmetric on the log scale, this crossing is an approximation, and the \texttt{EValue} output should be reported.

\subsection{One plotting function for four frameworks}

\begin{figure}[htbp]
\centering
\includegraphics[width=\textwidth]{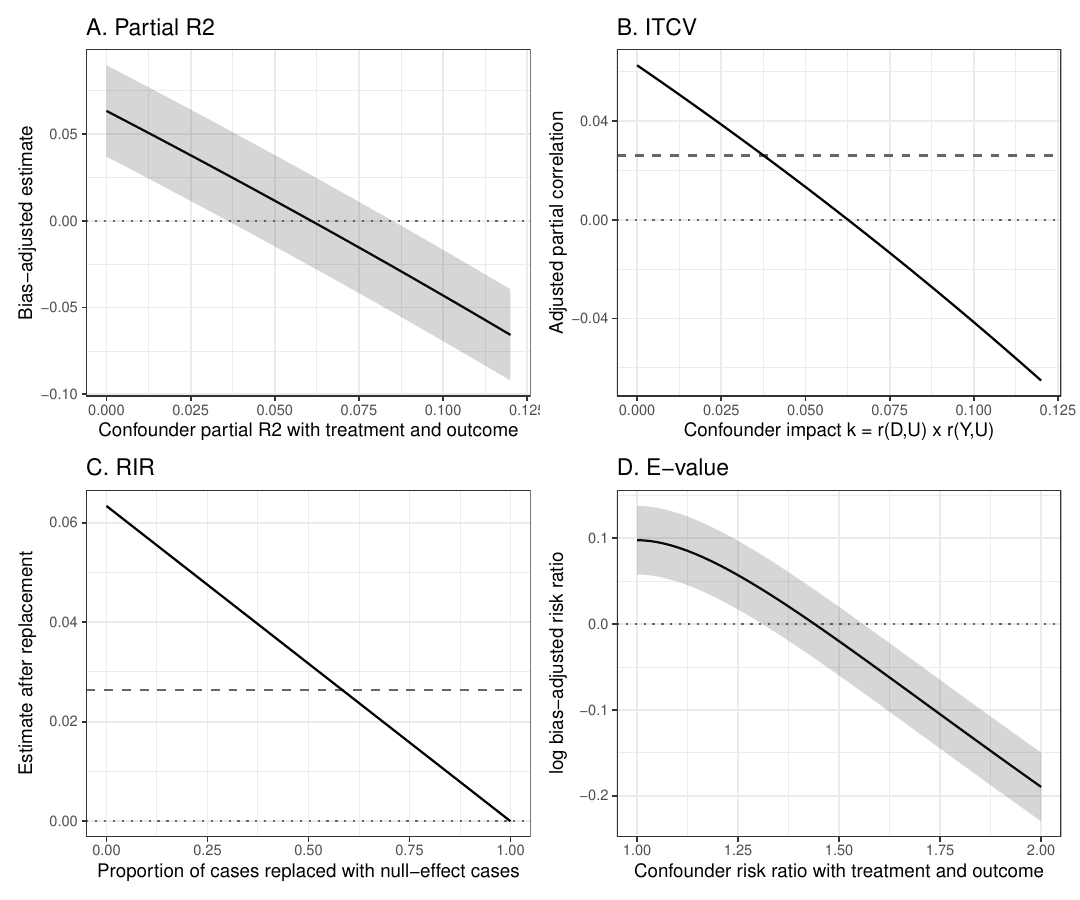}
\caption{Output of \fn{plot\_robustness\_curve()} for four frameworks on the RHC data. (A) Equal-strength partial-$R^2$ path with pointwise $\pm1.96$ adjusted-standard-error bands. (B) ITCV path; the dashed line is the critical correlation. (C) RIR path; the dashed line is $\beta^{\#}$. (D) E-value path on the log risk-ratio scale, with bands from the log-scale standard error reconstructed from the confidence interval. Dotted lines mark the null. The horizontal axes have different units and cannot be compared across panels.}
\label{fig:paths}
\end{figure}

Figure~\ref{fig:paths} applies \fn{plot\_robustness\_curve()} to the partial-$R^2$, ITCV, RIR, and E-value paths. The panels are combined with \texttt{patchwork} \citep{patchworkpkg}. Each panel takes its axis labels and reference lines from its own \fn{meta} list.

\par\noindent\begin{minipage}{\linewidth}
\begin{lstlisting}
library(ggplot2)
library(patchwork)
f1 <- plot_robustness_curve(p_r2, points = FALSE) + ggtitle("A. Partial R2") +
      plot_robustness_curve(it$path, points = FALSE) + ggtitle("B. ITCV") +
      plot_robustness_curve(rr$path, points = FALSE) + ggtitle("C. RIR") +
      plot_robustness_curve(p_ev, points = FALSE) + ggtitle("D. E-value") +
      plot_layout(ncol = 2) & theme_bw(base_size = 9)
\end{lstlisting}
\end{minipage}\par

\subsection{Measured-covariate benchmarks}
\label{sec:bench}

\fn{covariate\_impacts()} evaluates every measured term of the model on either regression metric and attaches the threshold as an attribute.

\par\noindent\begin{minipage}{\linewidth}
\begin{lstlisting}
ci <- covariate_impacts(fit, treatment = "RHC", metric = "itcv")
cr <- covariate_impacts(fit, treatment = "RHC", metric = "partial_r2")
head(ci[order(-abs(ci$impact)), c("covariate", "r_du", "r_yu", "impact")], 3)
head(cr[order(-cr$impact), c("covariate", "r2dz.x", "r2yz.dx", "impact")], 3)
\end{lstlisting}
\end{minipage}\par
\par\noindent\begin{minipage}{\linewidth}
\begin{lstlisting}[style=Rout]
   covariate        r_du        r_yu       impact
16  surv2md1 -0.08834258 -0.10178910  0.008992311
19   wtkilo1  0.08423773 -0.06965977 -0.005867981
64  dnr1_Yes -0.06639700  0.06598209 -0.004381013
   covariate      r2dz.x     r2yz.dx      impact
16  surv2md1 0.007804411 0.009240645 0.008525553
19   wtkilo1 0.007095995 0.005594437 0.006323117
64  dnr1_Yes 0.004408562 0.004914740 0.004665070
\end{lstlisting}
\end{minipage}\par

\begin{figure}[htbp]
\centering
\includegraphics[width=\textwidth]{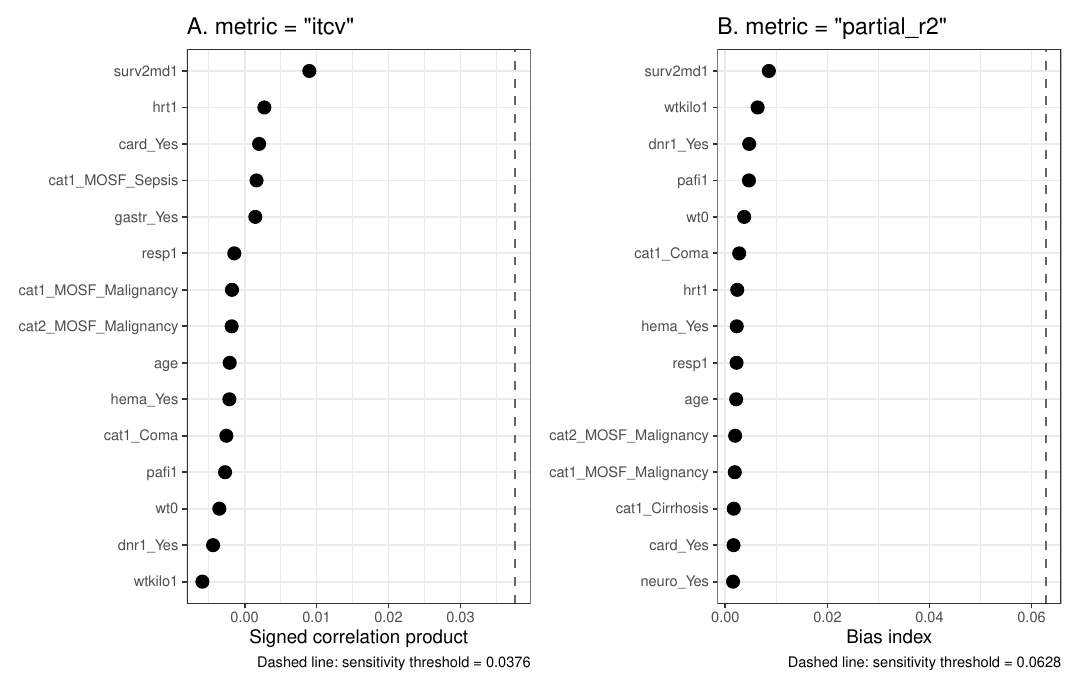}
\caption{Sensitivity Love plots from the two outputs of \fn{covariate\_impacts()}. (A) Signed ITCV products; positive values act to reduce the estimate, and the dashed line is the conditional ITCV. (B) Partial-$R^2$ bias indices; the dashed line is $|t|/\sqrt d$. The panels use different metrics and thresholds.}
\label{fig:love}
\end{figure}

The two-month survival estimate has the largest adverse ITCV product (0.0090) and the largest bias index (0.0085). Neither reaches its threshold: the ITCV of 0.0376, which marks the loss of significance, or the value $|t|/\sqrt d=0.0628$, which marks the point null. The other leading ITCV products are negative, which means that omitting those terms would act to strengthen the estimate. These comparisons are references, not bounds. A benchmark also does not decide whether a covariate belongs in the adjustment set; that requires causal knowledge about confounders, mediators, and colliders \citep{cinelli2024controls}. Figure~\ref{fig:love} shows the 15 terms with the largest absolute values on each metric.

\par\noindent\begin{minipage}{\linewidth}
\begin{lstlisting}
f2 <- (plot_sensitivity_love(ci, top = 15) +
         labs(title = "A. metric = \"itcv\"", x = "Signed correlation product")) +
      (plot_sensitivity_love(cr, top = 15) +
         labs(title = "B. metric = \"partial_r2\"", x = "Bias index")) &
      theme_bw(base_size = 9)
\end{lstlisting}
\end{minipage}\par

The Love plot shows only the product of the two correlations. \fn{plot\_sensitivity\_contour()} shows the two correlations themselves. Passing the \fn{itcv\_lm()} result draws the exact boundaries of Equation~\ref{eq:radj}.

\par\noindent\begin{minipage}{\linewidth}
\begin{lstlisting}
top3 <- head(ci[order(-abs(ci$impact)), ], 3)
f3 <- plot_sensitivity_contour(itcv = it, benchmarks = top3,
                               limits = c(-0.35, 0.35)) +
      theme_bw(base_size = 9)
\end{lstlisting}
\end{minipage}\par

\begin{figure}[htbp]
\centering
\includegraphics[width=.8\textwidth]{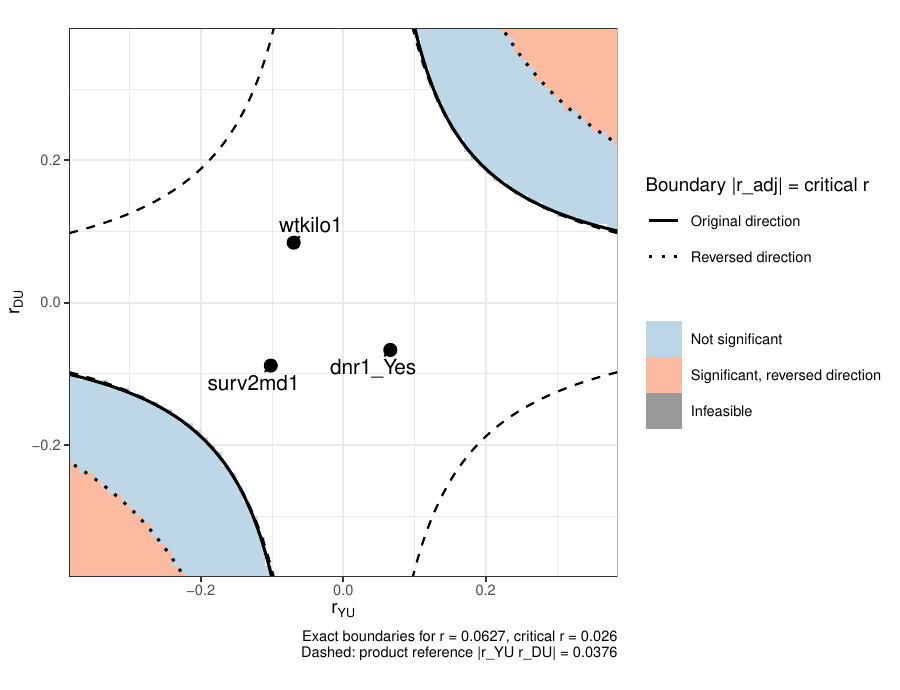}
\caption{Exact ITCV boundaries drawn by \fn{plot\_sensitivity\_contour()} for the RHC estimate, with the partial correlations of three measured terms. Solid curves: the adjusted correlation equals the critical correlation in the original direction. Dotted curves: the same in the reversed direction. Dashed curves: the product reference $|r_{YU}r_{DU}|=0.0376$. It meets the solid boundary only at the point where both correlations equal $\sqrt{0.0376}$ in the adverse direction; its branches in the other two quadrants correspond to confounders that strengthen the estimate. Unshaded: the conclusion holds. Here the horizontal axis is $r_{YU}$ and the vertical axis is $r_{DU}$.}
\label{fig:contour}
\end{figure}

Figure~\ref{fig:contour} shows that the product curve and the exact boundary meet at one point on the diagonal and separate elsewhere. Away from that point, a confounder with one strong and one weak correlation needs a product larger than the ITCV to remove significance, so the product curve is a conservative reference. All three measured terms lie well inside the unshaded region.

\subsection{Two-parameter surfaces}

\fn{sens\_surface()} returns the surface of which a path is one slice. Below we compute the partial-$R^2$ surface from the imported \texttt{sensemakr} result, so that its formal bounds are drawn automatically, and the E-value surface from the imported \texttt{EValue} result.

\par\noindent\begin{minipage}{\linewidth}
\begin{lstlisting}
ax <- seq(0, 0.1, length.out = 201)
s_r2 <- sens_surface(p_sm, a = ax, b = ax)
s_ev <- sens_surface(p_ev)
s_r2
f4 <- plot_sensitivity_surface(s_r2, benchmarks = head(cr[order(-cr$impact), ], 2)) +
        ggtitle("A. Partial R2 (sensemakr result)") +
      plot_sensitivity_surface(s_ev) + ggtitle("B. E-value") +
      plot_layout(ncol = 1) & theme_bw(base_size = 9)
\end{lstlisting}
\end{minipage}\par
\par\noindent\begin{minipage}{\linewidth}
\begin{lstlisting}[style=Rout]
<confoundsurf>
  framework : partial_r2
  a (x-axis): Partial R2 of U with treatment [0, 0.1]
  b (y-axis): Partial R2 of U with outcome [0, 0.1]
  value     : Adjusted estimate
  boundaries: Null (point estimate = 0); Significance (alpha = 0.05); Significance, reversed direction (alpha = 0.05)
\end{lstlisting}
\end{minipage}\par

\begin{figure}[htbp]
\centering
\includegraphics[width=.8\textwidth]{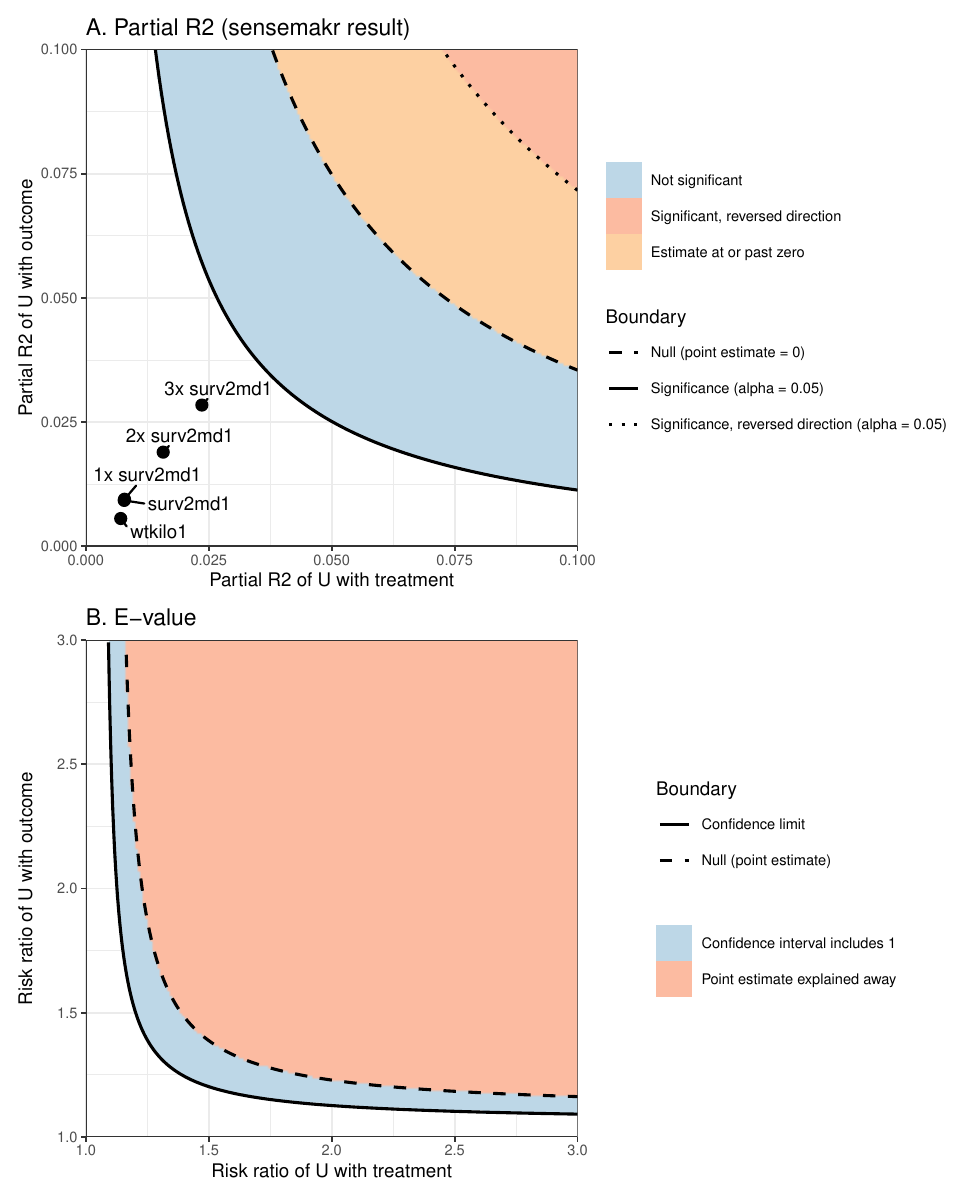}
\caption{Two-parameter surfaces drawn by \fn{plot\_sensitivity\_surface()}. (A) Partial-$R^2$ surface for the RHC estimate. Solid curve: significance is lost. Dashed curve: the point estimate reaches zero. Dotted curve: significance in the reversed direction. Blue: not significant; orange: the point estimate is at or past zero; red: significant in the reversed direction. Points: the two strongest measured terms and \texttt{sensemakr}'s bounds for a confounder one to three times as strong as \fn{surv2md1}. (B) E-value surface. Dashed curve: pairs of risk ratios that explain away the point estimate; it passes through (1.44, 1.44). Solid curve: pairs that move the confidence limit to 1; it passes through (1.31, 1.31). Unshaded regions: the conclusion holds. The size of a shaded region depends on the axis ranges and has no inferential meaning.}
\label{fig:surface}
\end{figure}

In Figure~\ref{fig:surface}A, the solid curve crosses the diagonal at the significance robustness value (0.036), and the dashed curve crosses it at the robustness value (0.061). Away from the diagonal, the surface shows the other confounders that would have the same effect. A confounder that explained only 1\% of the residual variance of RHC would need to explain about 14.7\% of the residual variance of death to remove significance, a point beyond the upper edge of the panel. Even a confounder three times as strong as the two-month survival estimate stays in the unshaded region. In Figure~\ref{fig:surface}B, the familiar E-value of 1.44 is the point where the dashed curve meets the diagonal. A confounder with a risk ratio of 3 with treatment would need a risk ratio of only 1.16 with death to explain away the point estimate.

\subsection{Several analyses in one table}

\fn{sens\_table()} collects the decision information of several paths. Each row keeps its own parameter, and the printed footer says so.

\par\noindent\begin{minipage}{\linewidth}
\begin{lstlisting}
tb <- sens_table(partial_r2 = p_r2, itcv = it$path, rir = rr$path, evalue = p_ev)
tb[, c("path", "summary_name", "summary_value", "lambda_null", "lambda_sig")]
\end{lstlisting}
\end{minipage}\par
\par\noindent\begin{minipage}{\linewidth}
\begin{lstlisting}[style=Rout]
<confoundvis_table>
 path       summary_name       summary_value lambda_null lambda_sig
 partial_r2 RV (point null)    6.083e-02     0.06083     0.03605
 itcv       ITCV (conditional) 3.758e-02     0.06265     0.03758
 rir        RIR (cases)        3.355e+03     1.00000     0.58490
 evalue     E-value (point)    1.439e+00     1.43900     1.31100

Each row keeps its own sensitivity parameter (column `parameter`); lambda values are comparable only within a row.
\end{lstlisting}
\end{minipage}\par

The print method rounds each value to four significant digits, so 1.43900 means 1.439. The table is intended for reporting several analyses side by side, not for ranking them. A robustness value of 0.061, an ITCV of 0.038, an RIR of 3,355 cases, and an E-value of 1.44 measure different things.

\FloatBarrier

\section{Worked example: matched pairs}
\label{sec:pairs}

The \texttt{werfel} data in \texttt{DOS2} come from \citet{werfel1998}. They compare a measure of DNA damage in 39 electric arc welders and 39 controls matched for age and smoking, and they are a standard example of Rosenbaum's sensitivity analysis \citep{rosenbaum2020}. \fn{sens\_path\_pairs()} takes the treated-minus-control differences.

\par\noindent\begin{minipage}{\linewidth}
\begin{lstlisting}
data("werfel", package = "DOS2")
d <- werfel$serpc_p - werfel$cerpc_p
p_gam <- sens_path_pairs(d, gamma = seq(1, 6, by = 0.01))
robustness_points(p_gam)
sens_report(p_gam)
f5 <- plot_robustness_curve(p_gam, points = FALSE) + labs(title = "A. Path", x = "Gamma") +
      plot_sensitivity_surface(sens_surface(p_gam)) + ggtitle("B. Surface") +
      plot_layout(ncol = 2, widths = c(1, 1.35)) & theme_bw(base_size = 9)
\end{lstlisting}
\end{minipage}\par
\par\noindent\begin{minipage}{\linewidth}
\begin{lstlisting}[style=Rout]
   level theta_start lambda_null lambda_sig        sig_rule
1 pooled    4.535374          NA    3.86761 theta threshold
Sensitivity report (framework: rosenbaum; source: pairs)
Matched pairs: 39 (39 with nonzero difference); Wilcoxon signed rank
  statistic = 715; sensitivity value Gamma = 3.87 (one-sided alpha =
  0.05).
Confounder strength (lambda): Gamma (odds of differential treatment
  assignment).
The upper-bound P-value reaches 0.05 at Gamma = 3.87; it stays below
  0.5 for Gamma <= 6.
These values describe how strong an omitted confounder would need to
  be. They do not show whether such a confounder exists, and they rest
  on the framework's own modelling assumptions. Judge plausibility
  against substantive knowledge and, where available, observed
  benchmarks.
\end{lstlisting}
\end{minipage}\par

The welders show more DNA damage. The result stays significant at $\alpha=0.05$ unless hidden bias could give one man in a pair 3.87 times the odds of the other of being the welder. The \fn{lambda\_null} entry is missing because the upper bound stays below 0.5 over the whole grid. The same \fn{robustness\_points()}, \fn{sens\_report()}, and \fn{plot\_robustness\_curve()} functions used for the regression paths apply without change.

\begin{figure}[htbp]
\centering
\includegraphics[width=\textwidth]{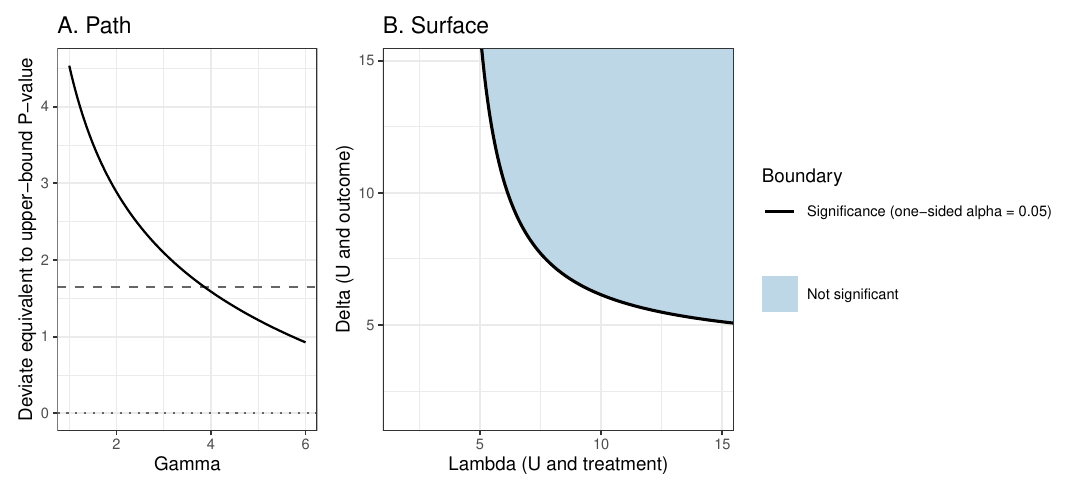}
\caption{Rosenbaum sensitivity analysis for the welder data. (A) Path drawn by \fn{plot\_robustness\_curve()}: the deviate equivalent to the upper-bound P-value; the dashed line is the one-sided critical value for $\alpha=0.05$, and the dotted line marks zero. (B) Surface drawn by \fn{plot\_sensitivity\_surface()}: each $(\Lambda,\Delta)$ pair on the curve gives $\Gamma=3.87$.}
\label{fig:pairs}
\end{figure}

Figure~\ref{fig:pairs}B gives the amplification of \citet{rosenbaum2009}. The single value $\Gamma=3.87$ corresponds to a curve of confounders. For example, a confounder that increased the odds of being a welder six-fold would need to increase the odds of a positive pair difference about 10.4-fold, and one with $\Lambda=10$ would need $\Delta\approx6.1$. These values agree with \texttt{sensitivitymv::amplify()}.

\section{Validation and reproducibility}
\label{sec:validation}

A software article should show that the package calculates what it claims to calculate. Three kinds of evidence are available for version 0.3.0.

First, the replication script checks the results against closed-form expressions and the upstream packages. Table~\ref{tab:validation} lists the checks. The script stops with an error if any check fails. Besides the two examples above, it uses three further reference cases: a constructed set of 162 binary matched pairs; the \texttt{darfur} data of \citet{hazlett2020}, which is the running example of \citet{cinelli2020}; and the worked RIR example of \citet{frank2025}.

\begin{table}[htbp]
\caption{Numerical checks in the replication script}
\label{tab:validation}
\centering\small
\setlength{\tabcolsep}{4pt}
\renewcommand{\arraystretch}{1.15}
\begin{tabularx}{\textwidth}{@{}p{.22\linewidth}Xp{.14\linewidth}@{}}
\toprule
\textbf{Data} & \textbf{Quantity and reference} & \textbf{Agreement} \\
\midrule
RHC & Robustness value vs.\ Equation~\ref{eq:rv} and \texttt{sensemakr} \fn{rv\_q} & $<10^{-12}$ \\
RHC & Significance crossing vs.\ \texttt{sensemakr} \fn{rv\_qa} & $<10^{-6}$ \\
RHC & Conditional and unconditional ITCV vs.\ \texttt{konfound} & $<10^{-12}$ \\
RHC & RIR cases; percent bias vs.\ \texttt{konfound} & exact; $<10^{-8}$ \\
RHC & Imported and native paths (partial $R^2$, ITCV, RIR) & identical \\
RHC & E-value crossings vs.\ \texttt{EValue} & grid step $10^{-3}$ \\
RHC & Partial-$R^2$ surface diagonal vs.\ \texttt{sensemakr} \fn{rv\_qa} & $<2\times10^{-3}$ \\
RHC & E-value surface null boundary vs.\ Equation~\ref{eq:bf} & $<10^{-10}$ \\
Welders & Wilcoxon upper-bound P-values vs.\ \texttt{DOS2::senWilcox()} & $<10^{-12}$ \\
Welders & Rosenbaum surface vs.\ \texttt{sensitivitymv::amplify()} & $<10^{-3}$ \\
Constructed pairs & McNemar bound vs.\ \texttt{sensitivity2x2xk::mh()} & $<10^{-10}$ \\
\texttt{darfur} & Robustness values vs.\ \texttt{sensemakr}; ITCV and RIR (677) vs.\ \texttt{konfound} & $<10^{-6}$; exact \\
\citet{frank2025} & RIR example: 2,603 cases, 41.19\% bias & exact \\
\bottomrule
\end{tabularx}
\end{table}

Second, the package source contains 83 \texttt{testthat} tests with 230 expectations. They cover the constructor, every entry route, the crossing, report, and table functions, the surfaces, and all plotting functions. When the upstream packages are installed, the tests compare results with them. Version 0.3.0 passed \texttt{R CMD check} with no errors or warnings on Windows, macOS, and Linux before its release on CRAN.

Third, one script reproduces the article. It contains every code chunk printed above in the same order, writes the printed output to a text file for comparison, saves Figures~\ref{fig:paths}--\ref{fig:pairs}, exports the numbers quoted in the text, and records session information.

These checks show that the package implements the published formulas and carries upstream results faithfully. They do not show that the sensitivity assumptions suit the RHC or welder applications. Agreement between implementations is not evidence that unmeasured confounding is absent.

\section{Discussion}
\label{sec:discussion}

We have seen that \texttt{confoundvis} operates as a common interface for established sensitivity analyses. A researcher can begin with a fitted linear model, matched-pair differences, an existing \texttt{sensemakr}, \texttt{konfound}, or \texttt{EValue} result, or a precomputed path. Once the result is represented as a \texttt{confoundsens} object, the package provides a common set of plotting, boundary-finding, reporting, and table functions. In the RHC example, this workflow produced a robustness value of 0.061, a conditional ITCV of 0.038, an RIR of 3,355 cases, and an E-value of 1.44 through one set of functions, while each number kept its own scale and source.

This organization is useful because sensitivity analyses are often reported as isolated numbers. A robustness value, ITCV, E-value, or $\Gamma$ is easier to interpret when the reader can see the path that leads to the boundary and, where appropriate, how the threshold compares with measured covariates. The surfaces add a second view. Each published number is one point on a curve of confounders with the same effect: the robustness value on the diagonal of the partial-$R^2$ plane, the E-value on the diagonal of the risk-ratio plane, and $\Gamma$ on the amplification curve. The original authors already describe these planes \citep{cinelli2020,ding2016,rosenbaum2009}. The package draws them in one way for every framework that defines one, with measured covariates and imported bounds placed on the same axes. This supports the advice, common to all of these frameworks, to read a threshold together with its decision target and with observed benchmarks rather than as an isolated number \citep{cinelli2020,vanderweele2020,frank2025}. However, a common software object does not create a common statistical scale. The analyst should therefore preserve the framework name, decision target, and effect scale when several results are reported in the same study. The package helps by printing the parameter with every result and by returning the decision rule in the \fn{sig\_rule} column.

The examples also show the intended division of labor between \texttt{confoundvis} and the packages on which it builds. The native functions calculate the regression-based and matched-pair results, and the validation shows that they agree with the source packages. The E-value calculation is performed by \texttt{EValue} and then converted for common plotting. The package does not attempt to replace the source packages or change their assumptions. Similarly, an imported \texttt{sensemakr} or \texttt{konfound} object should remain available when more detailed framework-specific output is needed.

We acknowledge several limitations. First, the native regression functions in version 0.3.0 are designed for fitted linear models. The package does not provide its own sensitivity analysis for logistic, survival, or cluster-robust models; the RHC linear probability model was used only because the native functions require it. Second, the native Rosenbaum analysis covers matched pairs with Wilcoxon's statistic or McNemar's statistic. Matched sets, other test statistics, and other designs remain the domain of the Rosenbaum packages. Third, measured-covariate benchmarks are implemented for the two regression metrics but not for E-values or matched pairs. Fourth, the plotted bands are pointwise, and grid crossings are found by interpolation. Finally, sensitivity analysis addresses only one threat to observational inference. It does not correct limited overlap, measurement error, selection, or model misspecification \citep{hernan2020}.

These limitations define the scope of the package rather than a reason to combine additional methods inside the current article. Future versions can add model support when the corresponding statistical assumptions and reporting rules are clear. For the present version, the objective is more modest: make established sensitivity results easier to visualize, benchmark, compare, and report without changing their meaning.

We hope \texttt{confoundvis} will help researchers move from a single robustness number to a more transparent sensitivity report. The package is available from CRAN under the GPL-3 license, and the source, vignette, and issue tracker are publicly available through its project repository.

\section*{Computational details}

The results in this article were obtained with R 4.3.3 \citep{rcore2024} on Ubuntu 24.04 using \texttt{confoundvis} 0.3.0, \texttt{ATbounds} 0.1.1, \texttt{sensemakr} 0.1.6, \texttt{konfound} 1.0.3, \texttt{EValue} 4.1.4, \texttt{DOS2} 0.5.2, \texttt{sensitivity2x2xk} 1.01, \texttt{sensitivitymv} 1.4.4, \texttt{sandwich} 3.1-0, \texttt{ggplot2} 3.4.4, \texttt{ggrepel} 0.9.5, and \texttt{patchwork} 1.2.0. The examples were also run on Windows 11 with R 4.6.0. The package is available at \url{https://CRAN.R-project.org/package=confoundvis}; the development repository, tagged at release \texttt{v0.3.0}, is \url{https://github.com/subirhait/confoundvis}. Replication materials will accompany the journal submission.

\FloatBarrier
\clearpage
\setlength{\bibsep}{5pt}
\bibliography{references}

\end{document}